# A Lightweight Approach of Human-Like Playtesting

Yan Zhao
Virginia Tech
Blacksburg, VA, USA
yanzhao@vt.edu

Weihao Zhang
Virginia Tech
Blacksburg, VA, USA
weihaoz@vt.edu

Enyi Tang
Nanjing University
Nanjing, Jiangsu, China
eytang@nju.edu.cn

Haipeng Cai
Washington State University
Pullman, WA, USA
haipeng.cai@wsu.edu

Xi Guo
University of Science and Technology Beijing
Beijing, China
xiguo@ustb.edu.cn

Na Meng
Virginia Tech
Blacksburg, VA, USA
nm8247@vt.edu

*Abstract*—A playtest is the process in which human testers are recruited to play video games and to reveal software bugs. Manual testing is expensive and time-consuming, especially when there are many mobile games to test and every software version requires for extensive testing before being released. Existing testing frameworks (e.g., Android Monkey) are limited because they adopt no domain knowledge to play games. Learning-based tools (e.g., Wuji) involve a huge amount of training data and computation before testing any game.

This paper presents LIT—our lightweight approach to generalize playtesting tactics from manual testing, and to adopt the generalized tactics to automate game testing. LIT consists of two phases. In Phase I, while a human plays an Android game app *G* for a short period of time (e.g., eight minutes), LIT records the user's actions (e.g., swipe) and the scene before each action. Based on the collected data, LIT generalizes a set of *context-aware, abstract playtesting tactics* which describe under what circumstances, what actions can be taken to play the game. In Phase II, LIT tests *G* based on the generalized tactics. Namely, given a randomly generated game scene, LIT searches match for the abstract context of any inferred tactic; if there is a match, LIT customizes the tactic and generates a feasible event to play the game. Our evaluation with nine games shows LIT to outperform two state-of-the-art tools. This implies that by automating playtest, LIT will significantly reduce manual testing and boost the quality of game apps.

## I. INTRODUCTION

In the video game industry, **playtesting** refers to the process of exposing a game to its intended audience, so as to reveal potential software flaws during the game prototyping, development, soft launch, or after release. Human testers can sign up at playtest platforms to play video games and earn money [4, 5, 18]. Meanwhile, the mobile gaming industry has been growing incredibly fast. According to Sensor Tower, the worldwide spending in games grew 12.8% across the App Store and Google Play in 2019 [21]. By the end of 2019, 45% of the global gaming revenue came directly from mobile games; among all mobile apps, mobile games account for 33% of all app downloads, 74% of consumer spend, and 10% of all time spent in-app [2]. The booming mobile game industry has led to a rapid growth in the game testing demand, although it is always expensive and time-consuming to hire human testers to manually play games.

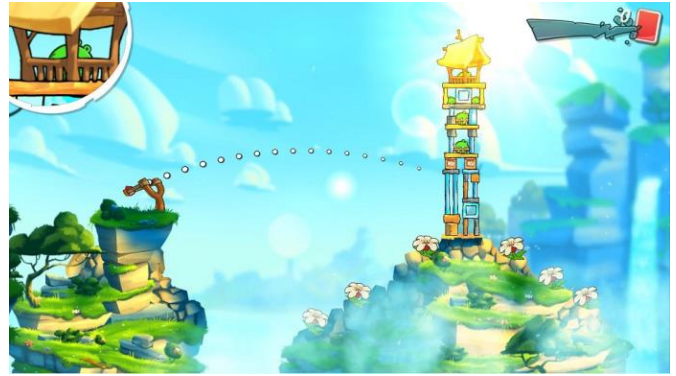

Fig. 1: A screenshot in the game *Angry Birds* [8]

Researchers and developers proposed approaches to automatically test Android apps and video games, but the tool support is insufficient. For instance, random testing (e.g., Android Monkey [15]) and model-based testing (e.g., AndroidRipper [22]) execute apps-to-test by generating various input events (e.g., button clicks) to trigger diverse program execution. However, these approaches only recognize the standard UI controls defined by Android (e.g., EditText, Button, and CheckBox [6]). They are unable to identify any customized playable UI items (e.g., the birds and pigs in the Angry Bird game as shown in Fig. 1); neither do they use any domain knowledge to effectively play games. Some approaches adopt machine learning techniques to test games by training a model with human-testing data [30, 45]. However, these approaches are heavyweight; they usually require for a large amount of training data, tremendous computational power, and careful hyper-parameter tuning before testing a game.

To help general developers efficiently test Android games without using complex ML techniques, in this paper, we present the design and implementation of a lightweight game testing approach—LIT. In our research, there are three challenges:

1) Games usually define various customized UI items or game icons (i.e., pictures) which are not recognizable by most automatic testing frameworks. To effectively play games, our approach needs to first identify those icons.
2) Different games define distinct rules and require users to play games by taking specialized actions (e.g., "long tap" or "swipe"). Our approach needs to mimic game-specific user actions to test games like a human.
3) A **game scene** is an image to display different information related to one program state (see Fig. 1). Scenes can be generated non-deterministically, so our approach should flexibly react to the changing program states.

To overcome the above-mentioned challenges, we developed LIT to have two phases: tactic generalization and tactic concretization. Here, a **tactic** describes in what context (i.e., program states), what playtest action(s) can be taken and how to take those actions.

The first phase requires users to (1) provide snapshots of game icons and (2) play the game *G* for a short period of time. Based on the provided snapshots, LIT leverages image recognition [16] to identify relevant icons in a given scene. When users play *G*, LIT recognizes each user action with respect to game icon(s) and further records a sequence of < *context, action* > pairs. Here, **context** removes scenery background but keeps all recognized game icons. From the recorded pairs, LIT generalizes tactics by (1) identifying abstract contexts $AC = \{ac_1, ac_2, \ldots\}$ and major action types $AT = \{at_1, at_2, \ldots\}$ and (2) calculating alternative parameters and/or functions to map each abstract context to an action type.

The second phase takes in any generalized tactics and plays *G* accordingly. Specifically, given a scene *s*, LIT extracts the context *c*, and tentatively matches *c* with any abstract context $ac \in AC$ involved in the tactics. If there is a context match, LIT randomly picks a corresponding parameter setting and/or synthesized function to create an action for game testing.

For evaluation, we applied LIT and two state-of-the-art testing tools (i.e., Monkey [15] and Sapienz [35]) to a set of game apps. Our evaluation shows that with an eight-minute user demonstration for each open-source game, LIT outperformed existing tools by achieving higher test coverage, passing more difficulty levels, and earning higher scores. Specifically for *CasseBonbons* [13] (an open-sourced version of the game *Candy Crush Saga* [12]), although the short user demo only passed 1 level and earned 4,050 points as the final score, LIT successfully passed 6 levels and earned 21,270 points, covering 77% of source code lines. On the other hand, Monkey could not even enter the game; it passed 0 level, earned 0 point, and achieved 4% line coverage. Sapienz could enter the game for playing; however, it only passed 0 level, earned 15 points, and achieved 50% line coverage. Among different runs of the same game, LIT worked stably well, while the performance of Monkey and Sapienz sometimes changed radically.

To sum up, we made the following contributions:
- We designed and implemented a novel algorithm to generalize tactics from user-provided icons and short game-playing demos. The algorithm identifies user actions, records < *scene, action* > pairs, and derives functions to map game context to feasible actions.

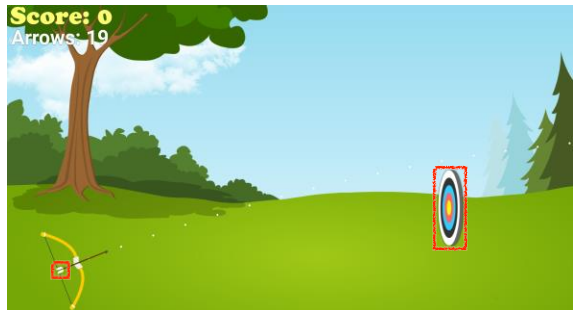

Fig. 2: A snapshot of the game *Archery*

- We designed and implemented a novel algorithm to test games based on the generalized tactics. LIT reacts to any randomly generated game scene by matching the scene with context in tactics, and taking actions accordingly.
- We compared LIT with two state-of-the-art tools using a dataset of nine games. Our evaluation comprehensively compared the test coverage of different tools in terms of source lines, branches, Java classes, and Java methods. We also compared the performance of different tools in terms of earned game scores, passed difficulty levels, and the stability. LIT outperformed both tools in all experiments.

## II. MOTIVATING EXAMPLE

This section uses an example to intuitively explain our research. *Archery* [10] is an open source Android game. As shown in Fig. 2, the game play is to shoot a target board with bow and a limited number of arrows in order to make a great score. The game is challenging in two ways. First, the target board will be placed randomly after each shot to the target. Second, if a user scores good, the board will move horizontally or vertically.

Suppose that a developer Alex wants to automatically test the game to reveal software flaws. As the board can be placed or moved randomly, naïve record-and-replay does not help automatic testing because the game scenes are generated in a non-deterministic way. Neither random testing nor model-based testing works because (1) arrow and board are game-specific graphic objects instead of standard GUI items, and (2) a user scores only if s/he pulls the arrow, shoots the arrow towards the board, and ensures the arrow to hit the board.

Our insight is that *when a user plays a game, user actions reflect the game-play tactics that are usable to automate game testing*. Thus, we designed LIT to work in two modes: **demo mode** and **testing mode**. Demo mode monitors how users play a game, and testing mode automatically tests the same game. To use LIT, Alex needs to provide two inputs: (i) snapshots of all relevant game icons and (ii) a game-play demo for a limited timespan. To prepare the first input, Alex can take a snapshot of the game and cut out images of board and arrow (see regions marked with **red rectangles** in Fig. 2). For the second input, Alex has to play the game in LIT's **demo mode**. Namely, once the game is launched, LIT takes a snapshot of the scene for record and prompts Alex to take an action as needed. While Alex swipes the arrow to complete an action based on his/her domain knowledge, LIT traces the finger. Afterwards, LIT takes another snapshot and prompts Alex for another action. This process continues until timeout.

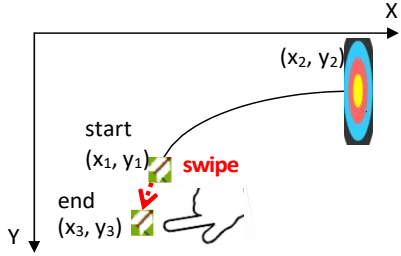

Fig. 3: Visualizing any $<context, action>$ pair for *Archery*

*Recognition of Context and Actions:* Based on the inputs and recorded data, LIT analyzes traces to identify Alex's action sequence and analyzes each scene snapshot to identify context. By indexing actions and context based on their timestamps, LIT creates a sequence of $<context, action>$ pairs $P = \{p_1, p_2, \ldots, p_n\}$. Intuitively, the information captured by a $<context, action>$ pair $p_i$ ($i \in [1, n]$) can be visually represented with Fig. 3. In the figure, every pixel of a display is represented with an xy-coordinate. The location of each game icon $o$ is represented with the coordinate of $o$'s centroid, such as $(x_1, y_1)$ for arrow and $(x_2, y_2)$ for board. The swipe operation is represented with a starting point $(x, y)$ and an ending point $(x_3, y_3)$, as indicated by the **red dotted directed edge**. Our goal of tactic inference is to generalize mappings from context to actions.

*Tactic Inference:* Based on recognized pairs, LIT analyzes three things for automatic testing:

- What is the commonality between context?
- What kind of actions are frequently applied?
- How is each context mapped to the corresponding action?

LIT infers the common context by comparing collected contexts, and finds the board and arrow to always exist while Alex plays the game. Similarly, LIT compares all identified actions and recognizes arrow-swiping as the major action type.

Each target-oriented swipe operation has three properties: (i) distance *dist*, (ii) direction *dir*, and (iii) duration *dur*. For simplicity, here we only explain the calculation of properties (i) and (ii) for any pair $p_i$. As shown in Fig. 4, LIT computes *dist* based on the coordinates of the start and end points. LIT calculates *dir* by synthesizing a function $f$ to cover coordinates of all three points, because such a function reflects Alex's angle to shoot the arrow. Namely, LIT first tries to synthesize a quadratic function $f_2(x) = ax^2 + bx + c$; if fails, LIT instead synthesizes a linear function $f_1(x) = kx + b$. Generally speaking, linear and quadratic functions are sufficient to cover any three given xy-coordinates. If $f_2$ is successfully synthesized, LIT records $a$ as the inferred direction because $a$ decides the width and direction (up or down) of a parabola's opening [20]. Similarly, if $f_1$ is created, LIT records $k$ as the inferred direction because $k$ decides the slope of $f_1$'s line. To sum up, LIT generates a tactic from Alex's inputs (see Table I).

*Tactic Application:* When testing *Archery*, given a randomly generated scene, LIT first identifies all game icons (i.e.,

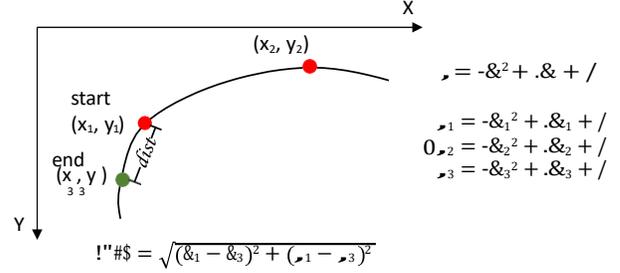

Fig. 4: Tactic inference from any $<context, action>$ pair for *Archery*

TABLE I: The tactic inferred from Alex's inputs

| "Abstract Context": | Actionable (arrow) Target (board) |
|---|---|
| "Action Type": | Swipe (actionable) |
| "Swipe Distance": | $dist_1, dist_2, \ldots, dist_n$ |
| "Swipe Direction": | Linear ($k_1, k_2, \ldots, k_l$) Quadratic ($a_1, a_2, \ldots, a_m$) |
| "Swipe Duration": | 0.26 (second), 1.26, ..., 0.33) |

arrow and board). To swipe an arrow towards the board, as shown in Fig. 5, LIT needs to decide the end point $(x', y')$ for the swipe operation. LIT randomly picks a distance $dist_i$, a direction parameter $p$, and a duration $dur_j$ from its tactic. If $p = k_j$, LIT solves the following equation group to get $(x', y')$:

$$(y' - y_1')^2 - (x' - x_1')^2 = dist_i^2 \quad (1)$$
$$y_1' - y' = k_j \times (x_1' - x_1') \quad (2)$$

Otherwise, if $p = a_l$, LIT solves the following equations:

$$(y' - y_1')^2 - (x' - x_1')^2 = dist_i^2 \quad (3)$$
$$y_1' - y' = a_l \times (x_2'^2 - x_2'^2) + b \times (x_1' - x') \quad (4)$$
$$y_2' - y_1' = a_l \times (x_2'^2 - x_2'^2) + b \times (x_2' - x_1') \quad (5)$$

Due to the random combination between inferred parameters, LIT does not guarantee all arrows to hit the board. However, all generated actions are valid arrow-shootings and some actions are highly likely to score. By diversifying the generated actions, LIT can test the game like humans and save Alex significant amount of time and manual effort.

## III. APPROACH

As illustrated in Fig. 6, there are two phases in LIT: tactic generalization and tactic concretization. In the following subsections (Section III-A–Section III-F), we will introduce the major steps in each phase.

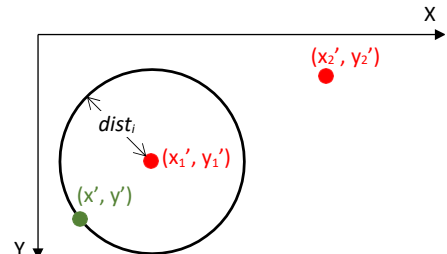

Fig. 5: Tactic application given a random scene of *Archery*

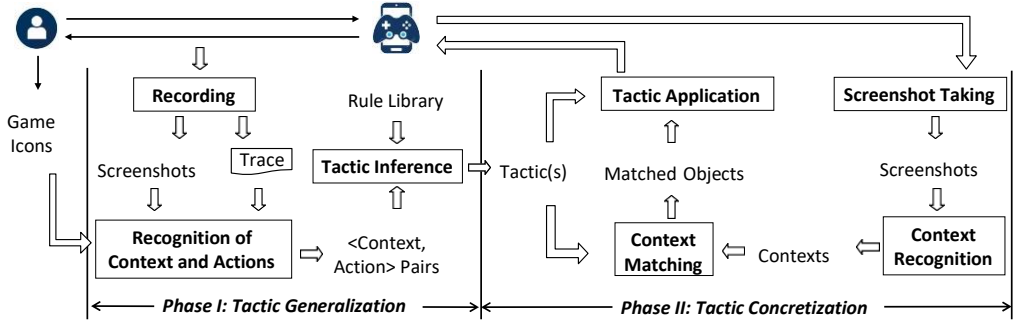
Fig. 6: LIT consists of two phases: tactic generalization and tactic concretization

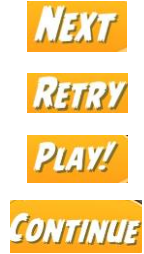

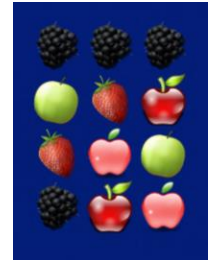

Fig. 7: An excerpt of a trace file

Fig. 8: Exemplar function icons in *Angry Birds*

Fig. 9: A screenshot of *AndroidLinkup* [7]

## A. Recording

To record the screenshots and traces while a user plays game $G$, we leveraged Android Debug Bridge (adb) [29]—a versatile command-line tool that enables developers to communicate with a device. The adb commands facilitate a variety of device actions, such as taking screenshots and recording event traces. When users play a game on phones or emulators, adb enables us to collect data and save data into our computer.

Here, one challenge to solve is: *How do we coordinate screenshot taking with the user's interactions with $G$?* On the one hand, we do not want to record screenshots too often, because these image files are large and too many generated files can quickly fill up our storage. On the other hand, we do not want to miss any scenes where a user takes actions, because the trace files needs to be later interpreted based on corresponding screenshots.

To tackle the challenge, we built LIT to periodically take screenshots and prompt the user to take actions. Specifically, in every nine seconds, LIT reads the system clock to acquire a timestamp $t$, takes a snapshot of the game screen, and names the picture file with "png.$t$". Depending on how complex a game scene is, LIT may spend 1–2 seconds generating the image file. Afterwards, LIT creates a trace file to record finger movements, and names that file with "txt.$t$". In a terminal, LIT then prompts the user "Please take an action to play the game". When the user takes an action based on the game-playing rule, LIT records all input events in the trace file. Such periodic operations continue until the demo period times out. In this way, screenshots and trace files can be aligned based on their common timestamps in file names. Each screenshot corresponds to exactly one trace file, and each trace file records data for exactly one user action.

Fig. 7 shows an excerpt of a trace file. In the file, the first column presents the timestamp of each event, although these timestamps cannot be mapped to the system-level timestamp $t$ mentioned above. All ABS_MT events report details on how an object (e.g., a finger) touches the screen and makes movements. Particularly, ABS_MT_POSITION_X and ABS_MT_POSITION_Y events show the xy-coordinates of contact points in a temporal order. When a finger moves on the screen, multiple xy-coordinates are recorded to reflect the trajectory.

## B. Recognition of Context and Actions

LIT recognizes context based on user-specified game icons. Currently, users are supported to specify three types of icons:

- *Actionable*—the icons that a user controls or plays with to score (e.g., arrow in *Archery*),
- *Target*—the icons that a user does not operate but are helpful for the user to decide how to operate actionable icons (e.g., board in *Archery*), and
- *Function*—the icons that a user manipulates to switch between major game phases, such as moving on to the next difficulty level or retrying the current level. Fig. 8 lists some function icons used in *Angry Birds*.

The user-specified categorized icons serve two purposes. First, they enable LIT to generalize *context-aware* tactics. If no icon is specified, LIT infers human testers' behavior patterns solely based on trace files. Second, if the user demo only involves a subset of specified icons, the category information allows LIT to generalize inferred tactics from involved icons to uninvolved ones. For instance, suppose that a demo only uses two of the four function icons shown in Fig. 8. LIT generalizes any tactic inferred based on these two icons to other same-type icons. Such approach design enables LIT to effectively infer tactics without requiring users to play games for quite a long time.

To recognize specified icons in given screenshots, we chose to use OpenCV (Open Source Computer Vision Library) [16] for image recognition. OpenCV is able to flexibly match images even if they are similar but different. Such flexibility is important for LIT to extract game icons from screenshots, because the specified icons are sometimes rotated, shadowed, or darkened in game scenarios. For each recognized image, OpenCV outputs coordinates of the matched area.

According to our experience, human testers play games with two major touch gestures: tap and swipe. Thus, in our research, we define a **user action** to include one or more touch gestures that are conducted for one valid move in games (e.g., shooting an arrow towards the board in *Archery*). Namely, a user action can be a single tap, a single swipe, or a number of gestures. In the *AndroidLinkup* game shown in Fig. 9, a user action consists of two taps, with each tap touching one fruit.

To recognize user actions, we implemented an intuitive approach to extract the gesture sequence from each trace file. With more details, we observed that the recorded event sequence for each touch gesture always (1) starts with ABS_MT_TRACKING_ID 0000xxxx, (2) ends with ABS_MT_TRACKING_ID 0000, and (3) has multiple ABS_MT_POSTION_X and ABS_MT_POSITION_Y events in between to show xy-coordinates of contact points. According to this observation, when processing each trace file, LIT first identifies one or more segments based on specific ABS_MT_TRACKING_ID formats. Inside each trace segment, suppose that the first xy-coordinate is $(x_f, y_f)$, the last xy-coordinate is $(x_l, y_l)$, and their related timestamps are separately $ts_f$ and $ts_l$. LIT then calculates two properties: distance $dist$ and duration $dur$.

$$dist = \sqrt{(x_l - x_f)^2 + (y_l - y_f)^2}$$

$$dur = ts_l - ts_f$$

With the two properties, LIT then derives a touch gesture from any segment based on the following two heuristics:

*H1:* If $dist > 20 \,\&\&\, dur > 0.2$ second, then a swipe gesture was made.

*H2:* If $dist < 20 \; dur < 0.2$ second, a tap gesture was made.

At the end of this step, LIT derives a sequence of < *context, action* > pair, with each pair corresponding to one recorded timestamp $t$.

### C. Tactic Inference

Given < *context, action* > pairs, LIT infers tactics by (1) identifying abstract contexts $AC = \{ac_1, ac_2, ...\}$ and major action types $AT = \{at_1, at_2, ...\}$ and (2) calculating alternative parameters and/or functions to map each abstract context to an action type. Namely, *each tactic consists of one abstract context, one action type, and a set of parameters and/or functions*.

To identify abstract contexts, LIT clusters collected contexts based on the number of icon types each context contains. For the *Angry Birds* game shown in Figure 1, some contexts include two icon types: actionable (i.e., birds) and target (i.e., pigs), while some contexts include only one icon type: function (i.e., "Next"). LIT considers each cluster to correspond to one abstract context $ac_i$, and represents $ac_i$ with the related icon types, as shown in Table I.

To identify major action types, LIT compares the actions related to each context cluster. If all or most of the actions are composed of the same gesture sequence $s$ (e.g., swipe), the inferred action type is also represented with $s$. Furthermore, in each of these < *context, action* > pairs, LIT tentatively maps the starting coordinate of action to game icons in the context; if the actions are always mapped to one icon type $i$, the inferred action type is refined to $s(i)$, as shown in Table I.

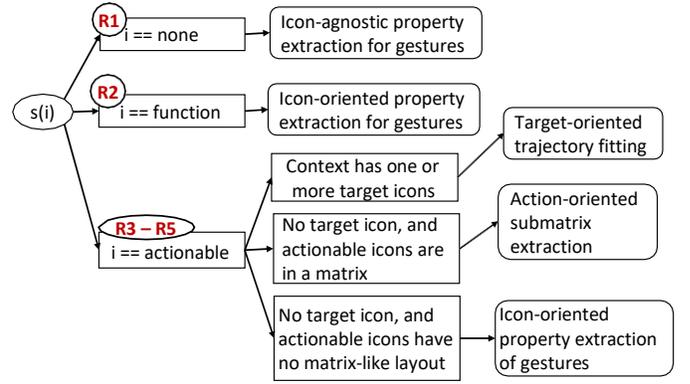

Fig. 10: Rules defined to infer parameters/functions for context-action mappings

The major challenge for this step is: *How do we calculate concrete parameters and/or functions to map each abstract context to an action type?* To overcome this challenge, given observed user actions and related contexts for each cluster, LIT follows the rules in our predefined library (see Fig. 10) to infer parameters and/or functions from each < *context, action* > pair. The inferred data describes given certain contexts, what concrete actions were taken by users. In later steps (Section III-E and Section III-F), LIT reuses such data to generate actions given a new context. Namely, the inferred data establishes concrete mappings from each abstract context to the related action type.

As shown in Fig. 10, the library currently has five rules. Given < *context, action* >, **R1** means that if an action was not applied to any specified icon, LIT extracts properties of individual gestures contained by the action. Particularly, for any tap, LIT extracts two properties: the starting coordinate $(x_f, y_f)$ and duration $dur$. For any swipe, LIT extracts three properties: distance $dist$, duration $dur$, and angle $sinx$.

$$sinx = \frac{y_l - y_f}{dist}$$

Similarly, **R2** describes that if an action was applied to a function icon, LIT extracts properties of individual gestures with respect to the icon. In particular, for any tap, LIT extracts one property—$dur$. For any swipe, LIT extracts three properties: $dist$, $dur$, and $sinx$.

**R3** describes the scenarios when an action was applied to an actionable icon and the context has one or more target icons (see Fig. 2). In such scenarios, according to our experience, users always made swipe gestures. Thus, LIT extracts above-mentioned three swipe properties for each gesture, and synthesizes a linear or quadratic function to fit a potential curve between the manipulated icon and a target. In the scenario when multiple target icons coexist, it is hard to guess at which target a user aims when manipulating an icon;

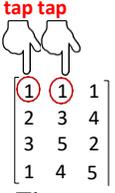

Fig. 11: The numeric representation of Fig. 9

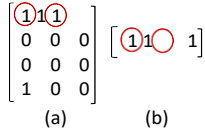

(a) (b)

Fig. 12: Normalized context and extracted submatrix

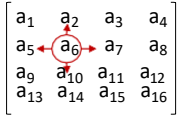

Fig. 13: Neighbors of a matrix element

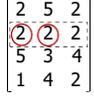

Fig. 14: LIT creates an action for a new context

thus, LIT randomly picks a target to synthesize the function. As illustrated by our motivating example, one coefficient of each synthesized function is saved for later use.

**R4** describes the scenarios when an action was applied to an actionable icon, and the context has no target icon but organizes all actionable icons in a matrix. Many match games [1] adopt such matrix layouts to place actionable icons. As shown in Fig. 9, the *AndroidLinkup* game lists different types of fruits in a matrix, and a user needs to tap two fruits of the same type to eliminate them both and earn score. If we use different numbers to refer to different fruit icons, a $<context, action>$ can be visualized as Fig. 11. We decided not to use such context as is in the inferred tactic for two reasons. First, randomly generated scenes can place fruits in arbitrary ways and the reusability of such context is quite limited in later steps. Second, not all elements in the matrix help explain the user action. Thus, we developed an action-oriented submatrix extraction algorithm to facilitate tactic inference and application.

**Algorithm 1** Action-oriented submatrix extraction

**Input:** context matrix $c$, matrix elements (i.e., actionable icons) involved in the action $E = \{e_1, e_2, \ldots\}$
**Output:** The extracted submatrix $sm$
1: **procedure** EXTRACTSUBMATRIX
2:    Queue q = ∅
3:    Initialize $sc = (minX, minY, maxX, maxY)$ to cover all elements in $E$
4:    Normalize $c$ to another matrix
5:    $q.enqueueAll(E)$
6:    **while** $q \neq ∅$ **do**
7:       $e = q.dequeue()$
8:       **if** any of $e$'s neighbor elements == "1" **then**
9:          Expand $sc$ to cover the neighbor element(s) as needed
10:      **end if**
11:    **end while**
12: **end procedure**

Based on our experience, icons in matrices are manipulated usually because they share certain commonality with surrounding context. Thus, we designed Algorithm 1 to extract an action-relevant submatrix (i.e., pattern) based on the commonality between manipulated elements and their neighborhoods.

In our algorithm, LIT first initializes a rectangle $sc$ to cover all elements in $E$. Secondly, LIT normalizes the given numeric matrix into a 0-1 matrix. Specifically, if an element is identical to any member $e \in E$, the element is converted to "1"; if the element is different from all members in $E$, it is converted to "0"; otherwise, if a grid in the matrix has no element, "-1" is used as a placeholder in the new matrix. For instance, Fig. 12 (a) presents the normalized representation for the matrix in Fig. 11. Thirdly, LIT enqueues all elements in $E$. For each dequeued element $e$, LIT checks individual neighbors (see Fig. 13) for the value "1". If a neighbor $n$ has "1", LIT enqueues $n$ and further examines whether $sc$ is large enough to cover $n$; if not, $sc$ is enlarged. This process continues until the queue is empty. Fig. 12 (b) shows the submatrix derived from Fig. 12 (a).

With $sc$ extracted, LIT infers a function $map(sc) = E$ from each $<context, action>$ pair. As what LIT does for **R2**, LIT also conducts icon-oriented property extraction for gestures. Therefore, the derived tactic includes $map$ functions and icon-related gesture properties.

**R5** describes the scenarios when an action was applied to an actionable icon, and the context has no target icon or matrix-like layout. Similar to what it does for **R2**, LIT simply extracts gesture properties with respect to the manipulated icons.

### D. Screenshot Taking And Context Recognition

This step reuses part of the implementation mentioned in Section III-A and Section III-B. Specifically, given game $G$, LIT periodically takes snapshots via adb, and relies on OpenCV and user-specified game icons to identify contexts.

### E. Context Matching

Given an identified context $c$, LIT tentatively matches $c$ with the abstract context of any derived tactic. If the icon types in $c$ exactly match that in $ac$, LIT concludes that $c$ matches $ac$.

### F. Tactic Application

Intuitively, this step is the opposite process of tactic inference. Namely, given a demo, tactic inference characterizes game contexts and derives a set of features to describe user actions. In comparison, this step leverages context characterization and derived features to randomly generate actions and relies on adb to issue those actions for game testing. Therefore, depending on what rule is adopted for tactic inference (see Section III-C), LIT applies tactics in distinct ways.

With more details, if **R1** is used for inference, LIT applies tactics by generating actions based on arbitrary property combinations between observed gestures. For instance, if a tap action is needed, LIT randomly picks a recorded coordinate $(x_f, y_f)$ and a duration $dur$ to create a tap. Similarly, if a swipe is needed, LIT creates the gesture by randomly picking $dist$, $dur$, and $sinx$ from its property sets. LIT similarly applies tactics if **R2** or **R5** is in use. When **R3** is used for tactic generalization, as illustrated by Section II, LIT randomly picks $dist$, direction parameter $p$, and $dur$ to decide how to swipe an actionable icon with respect to a target icon.

When **R4** is used for inference, to apply tactics to any given context $c$, LIT tentatively matches $c$ with any extracted submatrix $sc$. If there is a submatrix $sc'$ in $c$ such that (1) the elements matching 1's have the same icon index $i$ and

TABLE II: The nine Android games used in our evaluation

| Game | Type | LOC | Player's Actions | Context Characteristics |
|---|---|---|---|---|
| *Angry Birds* [8] | Close-source | - | Fling (or swipe) multiple colored birds to defeat green-colored pigs in a structure or tower. | With actionable icons (i.e., birds) and target icons (i.e., pigs) |
| *Ketchapp Basketball* [14] | Close-source | - | Swipe the ball towards the basketball hoop. | With actionable icons (i.e., balls) and a target icon (i.e., hoop) |
| *Star PopMagic* [19] | Close-source | - | Tap two or more adjacent identical stars to crush them. | With actionable icons (i.e., stars) organized in a matrix |
| *2048* [3] | Open-source | 1,692 | Swipe any point up/down/left/right to move the tiles. When two tiles with the same number touch, they merge into one. | Without actionable or target icon |
| *Apple Flinger* [9] | Open-source | 14,085 | Shoot (to swipe) apples towards enemy's base | With actionable icons (i.e., apples), but not organized in a matrix |
| *AndroidLinkup* [7] | Open-source | 2,102 | Tap two identical items to connect them with three or fewer line fragments and to crush them. | With actionable icons (i.e., fruits) organized in a matrix. |
| *Archery* [10] | Open-source | 2,833 | Shoot (or swipe) arrows towards a board. | With actionable icons (i.e., arrows) and a target icon (i.e., board) |
| *CasseBonbons* [13] (an open-source version of Candy Crush Saga [12]) | Open-source | 2,549 | Swipe colored pieces of candy on a game board to make a match of three or more of the same color. | With actionable icons (i.e., candies) organized in a matrix |
| *Open Flood* [17] | Open-source | 1,659 | Start in the upper left hand corner of the board. Tap the colored buttons along the bottom of the board to flood all adjacent filled cells with that color. | With actionable icons (i.e., buttons), but not organized in a matrix |

"-" means the data is unavailable.

(2) the elements matching 0's have indexes other than $i$, then LIT identifies elements for operation and creates an action by randomly mixing collected gesture properties. For instance, Fig. 14 presents a new context of *AndroidLinkup* that is totally different from the original context in Fig. 11 (a). When matching this context with the *sc* in Fig. 11 (b), LIT can locate two icons to tap and generate two taps accordingly.

## IV. EVALUATION

This section first introduces our data set (Section IV-A) and evaluation metrics (Section IV-B). It then explains how well LIT tests games (Section IV-C). Next, it expounds on the empirical comparison between LIT and two existing tools (Section IV-D and IV-E).

### A. Data Set

We included nine Android games into our evaluation set (see Table II); three of the games are closed-source commercial games and six games are open-source. We chose these games for two reasons. First, they are popular. Second, they present diverse context characteristics and require users to take various actions. With more details, users need to specify at least one function icon in each game so that LIT infers how to enter those games. Additionally, users need to specify actionable icons for some games (e.g., *CasseBonbons*), and specify both actionable and target icons for some other games (e.g., *Archery*). Each game requires users to either tap or swipe actionable icons. In Table II, column **LOC** shows the number of lines of code for each open-source game to describe the program size.

### B. Metrics

Similar to prior work [26, 44], we measured code coverage of execution by different testing tools to assess their effectiveness. Theoretically, the more program code is executed by a testing tool, the better the tool is. We defined four coverage metrics:

$$Line\_Coverage = \frac{\text{\# of lines of code covered}}{\text{Total \# of lines}} \times 100\%$$

$$Branch\_Coverage = \frac{\text{\# of code branches covered}}{\text{Total \# of branches}} \times 100\%$$

$$Method\_Coverage = \frac{\text{\# of methods covered}}{\text{Total \# of Java methods}} \times 100\%$$

$$Class\_Coverage = \frac{\text{\# of classes covered}}{\text{Total \# of Java classes}} \times 100\%$$

In our implementation, we adopted JaCoCo [27] to collect coverage information; and Jacoco uses the ASM library [11] to modify and generate Java byte code for instrumentation purpose. We computed the above-mentioned metrics based on the data collected with JaCoCo.

Actually, the above-mentioned coverage metrics are only computable for open-source games; they are not computable for close-source software because we have no access to the codebases. To also evaluate tools when they test close-source software, we defined two additional metrics: *Game_Score* and *Game_Level*. *Game_Score* reflects the points earned by a testing tool after it plays a game for a period of time; the higher score, the better. Similarly, *Game_Level* shows at which difficulty level a testing tool is when the allocated testing time expires; the higher level, the better.

### C. LIT's Effectiveness for Playtest

Given a game *G*, the first author manually played *G* for eight minutes in LIT's demo mode, and then switched to LIT's testing mode to automatically play *G* for one hour. Because there is randomness in the test inputs generated by LIT, we ran LIT to play each game five times such that each test run lasted for one hour. In this way, we evaluated not only how effectively LIT tests each game, but also how stable LIT's effectiveness is across different runs.

In Table III, columns **LIT** presents the average results of our tool across five runs, while columns **Demo** shows the results achieved by manual testing. In these four columns, "-" means that the data is not available. There are three reasons to explain such data vacancy in the table. First, some games do not show game scores (i.e., *AndroidLinkup* and *Open Flood*). Second,

TABLE III: The comparison of *Game Score* and *Game Level* among user demos, L<sub>IT</sub>, Monkey, and Sapienz

| Game | Game_Score | | | | Game_Level | | | |
|---|---|---|---|---|---|---|---|---|
| | Demo | L<sub>IT</sub> | Monkey | Sapienz | Demo | L<sub>IT</sub> | Monkey | Sapienz |
| *Angry Birds* | 179,394 | **1,147,827** | 35,546 | - | 2 | **7** | 0 | - |
| *Ketchapp Basketball* | 2 | **37** | 0 | - | 1 | **3** | 0 | - |
| *Star Pop Magic* | 695 | **2,805** | 225 | - | 1 | **2** | 1 | - |
| *2048* | 332 | **2,212** | 586 | 600 | - | **-** | - | - |
| *Apple Flinger* | 43,130 | **97,105** | 0 | 0 | 4 | **7** | 0 | 0 |
| *AndroidLinkup* | - | **-** | - | - | 2 | **5** | 0 | 1 |
| *Archery* | 180 | **493** | 0 | 0 | - | **-** | - | - |
| *CasseBonbons* | 4,050 | **21,270** | 0 | 15 | 2 | **7** | 0 | 1 |
| *Open Flood* | - | **-** | - | - | 1 | **6** | 0 | 1 |

"-" means the data is unavailable

TABLE IV: Code coverage comparison based on open-source games among L<sub>IT</sub>, Monkey, and Sapienz

| Game | *Line Coverage* (%) | | | *Branch Coverage* (%) | | | *Method Coverage* (%) | | | *Class Coverage* (%) | | |
|---|---|---|---|---|---|---|---|---|---|---|---|---|
| | L<sub>IT</sub> | Monkey | Sapienz | L<sub>IT</sub> | Monkey | Sapienz | L<sub>IT</sub> | Monkey | Sapienz | L<sub>IT</sub> | Monkey | Sapienz |
| *2048* | **81** | 80 | 77 | **68** | 65 | 62 | **84** | 86 | 84 | **78** | 78 | 78 |
| *Apple Flinger* | **54** | 19 | 9 | **53** | 17 | 7 | **61** | 27 | 16 | **60** | 32 | 24 |
| *AndroidLinkup* | **77** | 63 | 58 | **72** | 41 | 32 | **73** | 73 | 68 | **82** | 79 | 75 |
| *Archery* | **72** | 66 | 20 | **49** | 39 | 6 | **66** | 63 | 20 | **73** | 66 | 39 |
| *CasseBonbons* | **77** | 4 | 50 | **79** | 1 | 33 | **76** | 6 | 55 | **70** | 12 | 49 |
| *Open Flood* | **50** | 32 | 42 | **37** | 20 | 28 | **53** | 33 | 51 | **48** | 35 | 59 |
| Average | **68** | 44 | 43 | **60** | 30 | 28 | **69** | 48 | 49 | **69** | 50 | 54 |

some games have a single difficulty level instead of multiple levels (e.g., *Apple Flinger* and *Archery*). Third, Sapienz cannot execute with commercial games.

By comparing the **Demo** and **L<sub>IT</sub>** columns in Table III, we observed L<sub>IT</sub> to consistently outperform user demos by acquiring higher scores and passing more levels. For instance, in *Archery*, demo acquired 179,397 points and stopped at the $2^{nd}$ level; L<sub>IT</sub> obtained 1,147,827 points and stopped at the $7^{th}$ level. This means that L<sub>IT</sub> did not simply record or repeat what users did. Instead, it effectively inferred tactics from demos, and applied those tactics to react to randomly generated scenes. Our observation also indicates that with L<sub>IT</sub>, users do not need to manually test all games comprehensively. Instead, they can test the games for only a short period of time, and rely on L<sub>IT</sub> to spend more time testing the same games in similar ways.

The **L<sub>IT</sub>** columns in Table IV present the code coverage measurement for our tool. Among the six open-source games, L<sub>IT</sub> achieved 50–81% *Line_Coverage*, 37–79% *Branch_Coverage*, 53–84% *Method_Coverage*, and 48–82% *Class_Coverage*.

> **Finding 1:** *Based on eight-minute user demos, L<sub>IT</sub> effectively earned game scores, passed difficulty levels, executed lots of code within one-hour playtest.*

### D. Effectiveness Comparison with Existing Tools

To assess how well L<sub>IT</sub> compares with prior work, we also applied two state-of-the-art tools to our data set: Monkey [15] and Sapienz [35]. Two reasons explain why we chose these two tools for empirical comparison. First, Choudhary et al. [26] recently conducted an empirical study by running multiple automatic testing tools on the same Android apps, and revealed that Monkey outperformed the other tools in terms of code coverage and runtime overhead. Second, Mao et al. [35] conducted a more recent study and showed that Sapienz worked even better than Monkey. We believe that if L<sub>IT</sub> works better than both tools, it probably outperforms other tools as well. For fair comparison, we ran each of the chosen tools to test every game five times, with each run lasting for one hour; then we reported the average numbers across runs.

*a) Comparison Based on Game Score and Game Level:* As shown in Table III, L<sub>IT</sub> outperformed Monkey and Sapienz by always acquiring higher scores and passing more levels. For instance, when testing *Angry Birds*, L<sub>IT</sub> obtained 1,147,827 points and arrived at Level 7 with one-hour playtest. On the other hand, Monkey only got 35,546 points and did not pass any level within one-hour testing.

In two open-source games (i.e., *Apple Flinger* and *Archery*), neither Monkey nor Sapienz earned any point or passed any level. Two reasons can explain this phenomenon. First, the tools did not know how to enter the games and spent lots of time clicking random pixels on the display until accidentally hitting the "Play" button. Second, both games require players to swipe certain icons to hit targets. Because neither Monkey nor Sapienz has such domain knowledge, they could not properly generate swipe actions for scoring. In comparison, L<sub>IT</sub> inferred domain knowledge from user demos, and applied the inferred knowledge to score well.

*b) Comparison Based on Coverage Metrics:* According to Table IV, among the four metrics, L<sub>IT</sub> achieved 60–69% average coverage. Meanwhile, Monkey obtained 30–50% average coverage, while Sapienz acquired 28–54%. In particular, L<sub>IT</sub> always achieved higher coverage measurements than Monkey and Sapienz in four open-source games: *Apple Flinger*, *AndroidLinkup*, *Archery*, and *CasseBonbons*; it obtained comparable coverage values in the other two games: *2048* and *Open Flood*.

Three reasons can explain the observation. First, *2048* and *Open Flood* have relatively simple context and require for simple tap gestures, so Monkey and Sapienz could smoothly test both games by randomly clicking pixels in game scenes. Second, the other four games have more complex contexts (e.g., by including target icons or organizing actionable icons in a matrix) and/or require for carefully planned gestures. LIT has the domain knowledge to recognize distinct contexts and to plan actions accordingly, while the other tools do not. Third, for *2048* and *Open Flood*, Monkey and Sapienz not only tested the game-playing logic, but also tentatively explored other features like game configurations. As LIT focused on playing games, it earned higher scores and passed more levels but did not necessarily test more classes or methods.

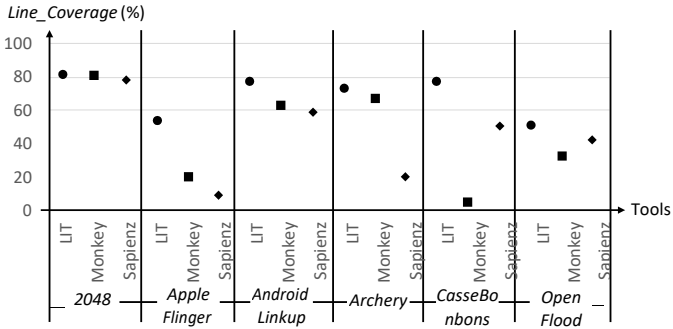

Fig. 15: The average-max-min chart of *Line_Coverage* by different tools

> **Finding 2:** LIT *outperformed Monkey and Sapienz by playing games more intelligently; it worked considerably better especially when game contexts are complex or actions need to be carefully planned.*

### E. Stableness Comparison with Existing Tools

When testing games via automatic tools, we expect the tools' coverage measurements to be stable across different test runs. This is because if a tool's effectiveness varies a lot among distinct runs, we need to run the same tool with the same game multiple times so that the best coverage is achieved in at least one run. We believe that such stability of tools' effectiveness affects their usability. Namely, the more stably a tool works, the more usable the tool is. Therefore, we also evaluated how stably each tool worked among different test runs.

Based on the five runs of each game by every tool, we plotted the average-max-min chart of *Line_Coverage* in Fig. 15. An average-max-min chart displays the average, maximum, and minimum values. In the chart, each node *n* represents an average value. The whisker going though *n* usually has two ends, representing the maximum and minimum values. Intuitively, the longer whisker a node has, the more variance there is between the measured five values.

According to Fig. 15, LIT outperformed Monkey and Sapienz in two aspects. First, LIT worked pretty stably across runs. Meanwhile, Sapienz worked less stably because it has short whiskers in all games. The stability of Monkey varies a lot across games: it worked stably in *AndroidLinkup* and *CasseBonbons*, was less stable in *2048* and *Archery*, and was least stable in *Apple Flinger* and *Open Flood*. Second, LIT outperformed the other tools in almost all runs. In five games (except *2048*), LIT's coverage values are universally better than the highest values of the other tools. The three tools worked very similarly only in game *2048*, mainly because the game is relatively simple. By randomly clicking pixels on the display, a tool can usually play the game smoothly.

> **Finding 3:** LIT *worked stably better than Monkey and Sapienz by acquiring higher Line_Coverage values in almost all runs of all games.*

## V. RELATED WORK

The related work of our research includes automated testing for Android apps, empirical studies on automated testing for Android apps, and automated game testing.

### A. Automated Testing for Android Apps

Various tools were proposed to automate testing for Android apps [15, 22, 23, 25, 28, 31, 32, 33, 34, 35, 38].

Specifically, random-based approaches (e.g., DynoDroid [33] and Monkey [15]) tests app by generating random UI events and system events. Given an app to test, model-based approaches (e.g., GUIRipper [22] and PUMA [31]) uses static or dynamic program analysis to build a model for the app as a finite state machine (FSM). An FSM represents activities as states and models events as transitions. The built model is then used to generate events and systematically explore program behaviors.

Since random-based and model-based approaches cannot trigger certain program behaviors that require for specific inputs, systematic exploration approaches (e.g., ACTEve [23] and EvoDroid [34]) were proposed to reveal such hard-to-trigger behaviors in order to increase testing coverage. In particular, ACTEve is a concolic-testing tool that symbolically tracks events from the point where they originate to the point where they are handled, infers path constraints accordingly, and generates test inputs based on the inferred constraints. However, these approaches do not recognize customized UI items or game icons, neither do they observe domain-specific rules to test games.

Record-and-replay tools (e.g., RERAN [28] and Mobi-Play [38]) record inputs and program execution when users manually test apps, and then replay the recorded data to automatically repeat the testing scripts. The record-and-replay methodology assumes that GUIs are always organized in a deterministic way and UI items are always put at fixed locations. However, when game scenes are randomly generated and game icons move, the above-mentioned assumptions do not hold.

Humanoid [32] is closely relevant to LIT. It leverages deep learning to infer how human users choose actions based on an app's GUI from human interaction traces, and then adopts the learned model to guide test input generation. The intuition is that Humanoid first trains a model with recorded traces from lots of existing apps. To test a new app *A*, Humanoid generates input events based on (1) *A*'s similarity with existing apps and (2) the frequent actions users take given similar GUIs. However, Humanoid does not test games; it is insensitive to any app-specific interaction modes because the trained model focuses on the commonalities between apps.

*B. Empirical Studies on Automated Testing for Android Apps*

Researchers conducted studies on automated testing for Android apps [26, 36, 37, 42, 43]. Specifically, Choudhary et al. [26] conducted a comparative study of main test input generation techniques for Android. Among the seven tools investigated, the researchers found Monkey [15] to achieve the highest testing coverage. Based on the empirical finding, Zeng et al. [43] applied Monkey to WeChat—a popularly used Android app, and revealed two limitations of Monkey. First, Monkey generated lots of redundant events. Second, Monkey is oblivious to the locations of widgets (e.g., buttons) and GUI states. Mohammed et al. [36] recruited eight users to test five Android apps, and simultaneously applied Monkey to test the same apps. They revealed that Monkey could mimic human behaviors when apps have UIs full of clickable widgets to trigger logically independent events. However, Monkey was insufficient to test apps that require information comprehension and problem-solving skills like games.

Our research was inspired by prior work. Some of our observations and experience corroborate prior findings.

*C. Automated Game Testing*

Several approaches were introduced to automate game testing [24, 30, 39, 40, 41, 45].

Specifically, online testing (e.g., TorX [39] and Spec Explorer [40]) is a form of model-based testing. With online testing, testers use a specification (or model) of the system's behavior to guide testing and to detect the discrepancies between the implementation under test (IUT) and the model. Both the IUT and the model are viewed as interface automata to establish a formal conformance relation between them. However, these testing methods require users to leverage formal specification languages to prescribe models. In comparison, LIT does not require users to learn or use any domain-specific language. Instead, LIT infers playtest tactics from user demos and uses the tactics to automate testing.

Deep learning-based approaches train models with lots of player data and use the models to predict the most "human-like" action in a given game scene [24, 30, 45]. For instance, Wuji [45] is the state-of-the-art tool that leverages evolutionary algorithms, deep reinforcement learning (DRL), and multi-objective optimizations to perform automatic game testing. When testing a game, Wuji intends to balance between winning the game and exploring the space. Since Wuji is not available even though we contacted the authors, we could not compare LIT with it empirically. These learning-based approaches usually require users (1) to provide a large amount of player data and lots of computing resources for game-specific training, and (2) to manually tune hyper-parameters for best performance. Developers may not afford the time, resource, and effort required for adopting these techniques, while the tools may not be able to test small Android games efficiently and effectively.

VI. THREATS TO VALIDITY

*Threats to External Validity:* All inferred tactics and empirical findings mentioned in this paper are limited to our experiment data set. Our rule library defined for tactic inference currently focuses on three major types of games: (1) games that require for no specialized consideration for the context (e.g., *2048*), (2) target-oriented games (e.g., *Archery*), and (3) match games (e.g., *CasseBonbons*). To generalizes our research, in the future, we would like to include more games into our evaluations. We will also include more inference rules into the library such that LIT can handle more diverse playtest tactics.

*Threats to Internal Validity:* When testing a game, LIT generates inputs by randomly combining gesture properties inferred from the user demo. Therefore, the quality of generated inputs is influenced by the user actions extracted from the demo. Namely, if a user takes many invalid actions in a demo (e.g., tapping a lot while swipes are expected), LIT is likely to conduct such invalid actions in testing; if a user purely takes valid and effective actions in a demo, LIT may also generate more valid actions for testing. To mitigate the effect of invalid user inputs, in the future, we can recruit more users to manually test the same game. By comparing the generated user demos, we can filter out invalid actions.

VII. CONCLUSION

As the mobile game market rapidly expands, more and more game apps become available and evolve quickly. There is an increasing demand for advanced testing methods to efficiently test games for better software quality and programmer productivity. Manual testing is expensive and time-consuming, while existing automatic tools are either too simple to test games or too complex for general developers to use. In this paper, we introduced a novel approach to achieve a good trade-off between the two factors of automatic testing: the testing effectiveness and the technical complexity.

Specifically, our new approach LIT requires users to specify game icons and to demonstrate how to play a game for a short period of time. LIT then infers playtest tactics from the demo and applies those tactics to automatically test the same game. Due to the feature of tactic inference, LIT's users do not need to learn or use any specification language to define the game model, neither do they have to provide lots of training data for machine learning. Thanks to the feature of tactic application, LIT's users can rely on it to autonomously decide how to respond to a randomly generated scene on-the-fly, and to generate appropriate actions based on that decision. Our evaluation shows exciting results of LIT; it also evidences the strength of rule-based tactic inference.

Currently, LIT currently includes five alternative inference rules in its library. These rules may be insufficient when LIT is used to test some complicated games (e.g., role-playing games). In the future, we would like to include more rules into the library to further improve LIT's applicability.